\documentclass[12pt]{article}
\usepackage{epsfig,subfigure,color,amsmath}
\usepackage{authordate1-4}

\begin{document}
\title{The bifurcation structure and noise assisted transitions in the Pleistocene glacial cycles}

\author{Peter D. Ditlevsen, \\
Centre for Ice and Climate, The Niels Bohr Institute,\\
University of Copenhagen, Juliane Maries Vej 30,\\
 DK-2100 Copenhagen O, Denmark.\\
}

\date{\today}
\maketitle

{\bf
The glacial cycles are attributed to the climatic response of the orbital changes in the 
irradiance to the Earth \cite{milankovitch:1930,hays:1976}. 
These changes in the forcing are to small to explain the observed climate variations as simple
linear responses. Non-linear amplifications are necessary to account for the glacial cycles.
Here an empirical model of the non-linear response is presented. From the model it is possible
to assess the role of stochastic noise in comparison to the deterministic
orbital forcing of the ice ages. The model is based on the bifurcation structure
derived from the climate history. It indicates the dynamical origin of the 
Mid-Pleistocene transition (MPT) from the '41 kyr world' to the
'100 kyr world'. The dominant forcing in the latter is still the 41 kyr obliquity cycle, but the
bifurcation structure of the climate system is changed. The model indicates that 
transitions between glacial and interglacial climate are assisted by internal stochastic noise
in the period prior to the last five glacial cycles, while the last five cycles are
deterministic responses to the orbital forcing.}

\section{Introduction}
The dominant orbital periods in solar insolation  
is the 41 kyr obliquity cycle (tilt of rotational axis, determining the meridional gradient in insolation) 
and the precessional cycles (determining the season when Earth is closest to the Sun) 
which decompose into 19 kyr and 23 kyr 
periods. However, through 
the last 800 kyr -- 1 Myr the dominant period for the glacial 
cycles is approximately 100 kyr similar to the one order of magnitude weaker 
eccentricity cycle (determining the semi-annual difference in distance to the sun). 
The weakness of this climatic forcing is referred to as the 
100 kyr problem of the Milankovitch theory \cite{imbrie:1993}.
It is now generally accepted that the 100 kyr glacial time scale cannot be attributed to 
the eccentricity cycle \cite{huybers:2007}. 
In the Plio- and early 
Pleistocene, 3--1 My BP, the dominant period of variation was indeed the 41 kyr 
obliquity variation \cite{raymo:1989,huybers:2005}. 
 
Different mechanisms have been proposed to explain the occurrence of the 100 kyr glacial cycle. 
These range from self-sustained non-linear oscillators \cite{kallen:1979,saltzman:1987,gildor:2000}, 
forced non-linear oscillators \cite{letreut:1983} to 
stochastic or coherence resonance \cite{benzi:1982,pelletier:2003}.
Others emphasize, based on
spectral analysis, the non-linear nature \cite{rial:1999} and the stochastic
nature of the climate signal \cite{kominz:1979,ashkenazy:2005}.

The non-linearity of the response to the forcing implies that a linear cross-spectral analysis 
between the paleoclimate record and the components of the orbital forcing does not necessarily 
pick out accurately the relative weights of the different components. 

Combined evidence from records of glaciations on land and deep sea 
records suggest that the climate has shifted between different 
quasi-stable states characterized by the mode of the global ocean circulation and the degree of 
glaciation \cite{imbrie:1992}. By comparison between the paleoclimatic record and the
non-linear stochastic model, it is demonstrated that the record can be generated by
the forcing from insolation changes due to the obliquity cycle through the
full record including the last 1 Myr.
It has been long known that the '100 kyr world' is not linearly responding
to the orbital forcing \cite{kominz:1979}, but 
even in the '41 kyr world' the climate response to the orbital forcing
is non-linear \cite{ashkenazy:2004}. The assumption here is that the orbital forcing
resulted in periodic jumps between
two stable climate states. 
What happened approximately 800 kyr -- 1 Myr ago was that a third deep glacial state became
accessible resulting in a change in length of the glacial cycles. 
The reason for this mid-Pleistocene transition (MPT) is unknown, and attributed to a 
gradual cooling due to a decreasing atmospheric $CO_2$ level \cite{saltzman:1991} or
a change in the bedrock erosion (the regolith hypothesis) \cite{clark:1998}. For a
review see \cite{clark:2006}.

For a simple 0-dimensional model of the Earth, where the climate is characterized by one number, the global mean 
temperature, there is ambiguity in ascribing the orbital forcing from the time and space varying insolation field 
across the globe.    

In a very interesting series
of papers Huybers et al. \cite{huybers:2005,huybers:2008} argue that integrated summer insolation 
is the relevant measure of orbital forcing. This is closely related to the concept of degree days, which
is the annual number of days with temperatures above freezing. The underlying assumption being that ablation from the
glaciers is a dominant climate driver. This measure is dominated by the obliquity cycle since, as noted by Huybers,
the increased insolation when Earth is close to the sun in its orbit is compensated by shorter time spend there
due to Keplers second law. Thus the total insolation during the degree days  becomes independent on the precessional cycle.   
Here we shall assume the integrated insolation exceeding 200 W/m$^2$ to be a proxy for the degree days forcing.

In contrast to this Paillard \shortcite{paillard:1998} shows, using a simple rule based model of jumping
between three different quasi-stationary climate states, that the climate record can be 
a response to the summer solstice insolation at 65N. The two proposed forcings (degree day insolation and
June 26 insolation) are different, since the latter has a strong component of the precessional cycle. 
Using the June 26 insolation as the better proxy for the forcing can be rationalized from the point of view
of a threshold crossing dynamics, since the extremal values (mid-summer insolation) would then be the governing parameter.   
However, since we cannot decide between the two within the framework of a simple model, we shall take
the alternative approach of assuming the linear combination of the two, considered as a first order expansion, 
which gives the best fit between
the observed record as response to the forcing.     

The starting point for the model is the rule based model proposed 
by Paillard. The rules imposed by Palliard will be
derived from an effective governing stochastic differential equation. This gives a dynamical
explanation of the rule based model and
potentially narrow the range of possible climate mechanisms and models capable
of explaining the glacial cycles.

\section{Non-linear climate response to the forcing}
Due to the high dimensionality and the stochastic nature of the climate fluctuations
it is highly unlikely that regular 
periodicities can result from internal oscillatory modes alone. 
It is much more plausible that non-linear responses to weak external periodic 
forcing would lead to periodic behaviour.  
There is evidence from observations as well as models that multiple
states exist in the climate system \cite{imbrie:1992}. This suggests a possible scenario of periodically induced
destabilizations of quasi-stable climate states.  


The multiple states governed by ice-albedo feedback as proposed in simple
energy balance models \cite{sellers:1969,budyko:1969}  have been
demonstrated in a more realistic climate model \cite{langen:2004}.   
Likewise it has been demonstrated that more realistic ocean models \cite{rahmstorf:1995} has 
a structure of stability and bifurcation points similar to the simple box models \cite{stommel:1961}. It is 
thus plausible that within the high dimensional climate system there are slow 
manifolds for which bifurcation points exist \cite{broecker:1997}.    

Bifurcation points in the system describe structural changes in the 
stability of meta-stable states as a function of a control parameter measuring the forcing of 
the system. The validity of linear analysis near the stable states permits a
complete classification of the possible types of bifurcation points in any 
non-linear system \cite{guckenheimer:1986}. 

The paleoclimatic records, especially from deep sea cores and ice cores, show
that at least three distinct climate states have been identified \cite{imbrie:1992}. These are: 
(1) A deep glacial state for which the North American and Fenno-Scandian ice sheets had 
their full extend and the Northern Atlantic deep water formation was weakened due to the 
extended sea ice cover. 
However, more recent analysis of sediment records show that even in the deep
glacial state there was a substantial thermohaline flow in the North Atlantic \cite{mcmanus:2004} 
(2) An intermediate or pre-glacial state where
the ice sheets were in a slow build up phase and the Gulf Stream made way for 
some intermediate water formation and the boreal heat pump. 
(3) The interglacial state where
the northern ice-sheets, except from Greenland, were gone and the deep water 
formation occurred as today north of Denmark straight with the Nordic heat pump. 
The fourth state described by Imbrie et al. \shortcite{imbrie:1992} 
will be regarded as a transition 
state here. 

The three states can be identified with the ones labelled 
$G$ (deep glacial), $g$ (pre-glacial), and $i$ (interglacial), respectively, 
by Paillard \shortcite{paillard:1998}.
Here we adapt the same notation. Paillard observed in the paleoclimatic
record that there seems to be "forbitten" transitions between the three states.
In the period 2-1 Myr BP the record shows regular oscillations between only
the two states $i$ and $g$, while in the period 1-0 Myr BP there is only a specific
sequence of occurences: $i \rightarrow g \rightarrow G \rightarrow i$ permitted.
The model presented here gives a dynamical explanation of this observation. 

\section{The stochastic model}
The model, which is empirical, assumes that 
the climate dynamics is reflected in a single
variable $x(t)$. This is as usual taken to be (minus) the global ice volume, represented by the deep sea
oxygen isotope ratio, roughly proportional to global 
mean surface temperature anomaly. The dynamics is described by an effective 
non-linear stochastic differential equation, 

\begin{equation}
dx=f_\alpha (x,\mu)dt + \sigma dB 
\end{equation}
where the white noise term $dB$ with intensity $\sigma$ describes the influence of the 
non-resolved variables and the internally generated chaotic climate fluctuations.
It is within this framework the roles of the orbital forcing and internal stochastic 
forcing is investigated. The deterministic part, $f_\alpha (x,\mu)$, of the dynamics depends on the external orbital
forcing, labelled by a single control parameter $\mu$ and internal parameters,
represented by $\alpha$. Note that Eq. 1 is non-autonomous, since $\mu$ and 
$\alpha$ are time dependent.

The full climate dynamics
can obviously not be completely reconstructed by such a single valued function. However, since stability and
bifurcations
are topological quantities it could be robust with respect to the detailed dynamics 
modelled. It is thus the bifurcation structure of $f_\alpha (x,\mu)$, 
with respect to the control parameter $\mu$, 
which determines the 
climate development. 

Guided by the observed record and
the transition rules proposed by Paillard we can empirically construct a bifurcation diagram:  
Figure 1, upper panel, shows the bifurcation diagram for the drift function $f_\alpha (x,\mu)$ as a
function of $\mu$ at the time interval 2 -- 1 Myr BP. 
The bifurcation diagram shows the curves $\{x_0(\mu)|f(x_0,\mu)=0\}$. The fat curves are the stable fixed point curves for 
which $\partial_x f < 0$, while the thin curves are the unstable fixed point curves for which $\partial_x f > 0$.
Thus in
the case of no additional noise ($\sigma=0$ in eq. 1) the state of the system $x(t)$ is uniquely determined from the
initial state $x(0)$ and the development of the forcing $\mu(t)$.
 
In the real climate system the internal noise is substantial and the system will not
reside exactly in the steady states determined by the bifurcation diagram. Thus the full drift function
needs to be parametrized.  The simplest way to parametrize the drift function in accordance with the
bifurcation diagram
is as a fifth order polynomial:

\begin{equation}f_\alpha (x,\mu)=\Pi_{j=1}^5(x-x^j_\alpha (\mu)),\label{poly}\end{equation}
where $x^j_\alpha (\mu)$ is the $j$'th steady state (zero-points) in the bifurcation diagram.
As labelled
in the figure the parameter $\alpha$ determines the position of the lower bifurcation point.
See the appendix for more details. A Matlab code of the model is available in the online supplementary material.  

It should be noted that this is, of course, not the only possible drift function corresponding to this
bifurcation diagram. In order to reconstruct the drift function from the observed realization, one could in principle
obtain the stationary probability density $p_{\mu_0}(x)$ by sorting $x(t)$ according to $\mu(t)=\mu_0$. Assuming 
that $\mu(t)$ is changing slowly in comparison to the time scale for $x(t)$ to drift to a stationary state $x_0$
($f_\alpha(x_0,\mu)=0$), one could then obtain $f_\alpha(x,\mu_0)$ by solving the Fokker-Planck equation \cite{gardiner:1985}
associated with Eq. 1 for fixed $\mu=\mu_0$. This would require a very long data series and complete
absense of additional non-climatic noise in the proxy data. This is not the case for the existing paleoclimatic record. 
 
The climate forcing is, as mentioned before, taken to be a linear combination of the summer solstice 65N insolation ($f_{ss}$) and 
the integrated summer insolation at 65N ($\overline{f}_I)$, where
the summer period is defined as the period where the daily mean insolation exceeds $I=200$ W/m$^2$.
The model results are robust with respect to the threshold $I$ chosen in a rather broad interval. 
The forcing, $f=\lambda \overline{f}_I+ (1-\lambda)f_{ss}$, shown in figure 3, second panel, is calculated 
using the code provided by Eisenman and Huybers \cite{huybers:2006}. Values of $\lambda$ around 0.5 gives the best
result, $\lambda=0.5$ is used. This assignment might, within the framework of the non-linear model, be interpreted as an
empirical determination of the dominating components of the orbital forcing.    

\section{The hysteresis behavior}
The diagram in figure 1, upper panel, shows the fixed points of $f_\alpha(x,\mu)$ as a function of
the deterministic forcing $\mu$. The three branches of stable fixed 
points $x^j(\mu)$ for the function, such that
$f_\alpha (x^j,\mu)=0$ and $\partial_x f_\alpha (x^j,\mu)<0$, are indicated by fat curves. 
The functional form for the five functions $x^j(\mu)$ are given in the appendix. The
specification of the $x^j(\mu)$'s and equations 1 and 2 completely defines the model.
Since $x$ is a proxy for global mean surface temperature anomaly (or minus global ice volume), 
the lower branch corresponds to the deep glacial state $G$. The middle branch corresponds 
to the climate state $g$ and the upper branch to the interglacial state $i$.  
The thin
curves correspond to the separating unstable fixed points. The dashed line-segments correspond to
pairs of complex conjugate roots in the fifth order polynomium. Note again that assuming a polynomial drift function, this is uniquely
determined from the roots, except from a trivial multiplicative constant.
 
Suppose now that the climate is in either of the states $g$ or $i$ and the climatic 
noise is too weak to induce a crossing of a barrier separating the stable states. Then
the only way a forcing induced shift between the climate states can occur is through 
bifurcations and a hysteresis as sketched by the arrows. Clearly the climate state $G$
is unreachable. 

Assume now that the lower bifurcation point, indicated by $\alpha_0$ in figure 1, upper panel, moves 
toward larger values of $\mu$ indicating that a stronger forcing is needed in 
order to destabilize the deep glacial state. 
In this case, $\alpha_0 \rightarrow \alpha_1$ shown in figure 1, lower panel, the glacial state $G$
is now reachable and a hysteresis loop $i$ $\rightarrow$ $g$ $\rightarrow$ $G$ 
$\rightarrow$ $i$ will appear. The central postulate of the model is the change in this
bifurcation structure represented by the shift of the point $\alpha$ (from $\alpha_0$ to
$\alpha_1$ on the $\mu$-axis)
at the Mid-Pleistocene transition.
This constitutes a dynamical explanation for the rule based model by Paillard \cite{paillard:1998}.
  
The change in the position of the lower bifurcation point is modelled such that 
$\alpha=\alpha_1$ when the climate is in state $i$. When the
state $G$ is reached through two bifurcations, $\alpha$ is gradually changing.
The gradual change in the bifurcation
diagram is modelled as a relaxation, $d\alpha/dt=-(\alpha-\alpha_0)/\tau$, where $\alpha_0$ is
the early Pleistocene equilibrium value and $\tau$ is a relaxation time.
When the climate bifurcates through 
the rapid transition $G \rightarrow i$, the parameter $\alpha$ again change to $\alpha_1$.

In order for the climate to skip the 41 kyr obliquity pacing of deglaciations the timescale $\tau$ governing
the bifurcation structure must be considerably longer than 41 kyr. The model results are quite insensitive to
the specific value of $\tau$ in the interval 70-130 kyr, and is set to be 100 kyr.
It is a major challenge to interpret the behavior of the bifurcation point $\alpha$, governed by
such a long time scale in terms of real climate dynamics. One could speculate that it is linked to the carbon
cycle and with erotion of continents on these long time scales.

\section{Comparison between the paleoclimatic record and the model}
The presence of the stochastic forcing implies that the climate
evolution is not fully deterministic. 

Figure 2, first panel, shows a 
particular realization of the model. The second panel shows the forcing, the red curve shows the value of $\alpha$, which is defined
as the position of the lower bifurcation point in figure 1 along the axis of the forcing (the x-axis). 
Note that a transition $G \rightarrow i$
without 
noise assistance is only possible when the forcing exeeds the value of alpha (that is
when the blue curve is above the red curve in the second panel).  

The composite 
Atlantic ocean  sedimentation record for the 
period 0 -- 2000 kyr BP  generated by Huybers \cite{huybers:2007} is shown in the bottom panel. 
The record is the benthic oxygen isotope sequence. 
The curve is plotted with normalized variance and the mean subtracted. This is
a proxy for the global ice volume.
The dating is based on a depth-age model independent from astronomical tuning. 
Note that the model resides predominantly in the deep glacial state $G$, rather than the 
pre-glacial state $g$. This is opposite to the model by Paillard \shortcite{paillard:1998}.

The differences between the single records compositing the stacked record gives an estimate
of the additional noise from bioturbation and other factors that makes the record different 
from a true record of ice volume. So
in order to compare the model with the observed climate record an additional red noise, of the same
magnitude as the difference in deep sea records is
added to the model. This is shown in the third panel, which should be compared with the observed record in fourth panel. 

\section{The role of the stochastic noise}
In order to investigate the role of the stochastic noise a set of realizations of the model has are presented
in figure 3. 
For the comparison between the model and the proxy climate records we focus on
the rapid transitions $G$ $\rightarrow$ $i$, called terminations \cite{broecker:1984,raymo:1997}. 
The top panel shows a realization with no stochastic noise. This is the deterministic climate response
to the orbital forcing. It is seen that the last five terminations are reproduced as observed, but there are fewer
interglacials in the earlier part of the late-Pleisocene period (1000-500 kyr BP) than in the observed record. From 
figure 2, second panel, it is seen that the amplitude of the orbital forcing is low in this period. The
bottom four panels show different realizations with a moderate stochastic forcing ($\sigma=0.8$ K/$\sqrt{kyr}$). In these 
realizations the added noise induces additional terminations at different times, suggesting a fundamental
unpredictability in glacial terminations.    

The transitions only happens with high probabilities at times when the orbital forcing is at a local 
maximum and when the value of the parameter $\alpha$ has become low enough (see figure 2, second panel). 
This will typically happen every 80 kyr or every 120 kyr, which we for now will denote short (S) and long (L) iceages.
A time period experiencing n short iceages and m long iceages is then denoted (nS,mL). There are 
then $K$(n+m,n) specific sequences (S,S,L,S,L,...) of iceage histories with n short and m long iceages, where $K$(n+m,n)=(n+m)!/(n!m!) is
the binomial coefficient. 
In the last 1 Ma ice age history there is the 
possibility of (11S,1L),(9S,2L),(8S,3L),(6S,4L),(5S,5L),(4S,6L),(2S,7L) and (0S,8L).
The number of possible iceage histories with only 80 kyr or 120 kyr durations 
is thus $$N=K(12,1)+K(11,2)+K(10,4)+K(10,5)+K(10,6)+K(9,2)+K(8,0)=941.$$  

The probabilities of transition are not equal for each maximum of the orbital forcing. It depends on two factors, the value
of the parameter $\alpha$, which depends on the time since the previous transition and the actual size of the forcing.
However,
assuming these to be equally likely, we
can roughly estimate the probability for the model to reproduce exactly the observed iceage history is only
around 1 permil. Note that this is a high estimate for the probability of the exact iceage history, since the
possibility of 40 kyr iceages is neglected. Such an iceage occured once in the model simulation shown in
figure 2. This compares well with the observed record (figure 2, lower panel) where the marine isotope stage 7
(approx. 250 -- 200 kyr BP) seems to be split into two periods.



Even though the observed iceage history is not expected to be reproduced in a given model realization,
the spectral density is similar.   
The spectral signature of the change in the climatic record from the 41 kyr world (2--1 Myr BP) to the 
100 kyr world (1--0 Myr BP) is shown
in figure 4.
The spectral signature is well reproduced in the randomly chosen model realization.

\section{Suggestive interpretation of the dynamics}
This section may be skipped without loss of continuity.
The model is empirically derived from the paleoclimatic record. Since the record is only a one 
dimensional representation of the climate state and the high dimensional real climate variability
is described as noise in an effective stochastic governing equation, the interpretation
of the underlying dynamics can only be suggestive. 

The climate stability can be described in terms of temperature-albedo feedbacks 
or equivalently ice volume-precipitation feedbacks  \cite{north:1981,ghil:1987,tziperman:2003}.
From the latter perspective we may write the non-linear
drift $f_\alpha (x,\mu)=$ ablation (ice melt off) - accumulation (precipitation on ice sheets). 
In figure 5 the bifurcation diagrams for the two periods are repeated (top panels) together
with graphs of the accumulation and ablation curves (a-f) along the transects shown in the
top panels. Going from top to bottom corresponds to increasing orbital forcing. 

The ablation, $f_\alpha^{abl}= c_0 + c_1\cdot (x-T_m)$, is assumed to be a linearly increasing function of the
temperature. The constants $c_0, c_1$ and the effective temperature of melting $T_m$ are related to the 
specific heat of melting, relation between global temperature and high latitude temperatures et cetera. It
could potentially be estimated from paleo-observations of freshwater influx into the oceans. 
The accumulation, $f_\alpha^{acc}$, is a more complex non-linear function 
of temperature. The accumulation is empirically determined from (2) simply as $f_\alpha^{acc}=
f_\alpha -f_\alpha^{abl}$. This splitting of the drift function should be considered 
schematic, and attempt towards more realism would involve much more complex climate models, which
shall not be pursued here.       
The stable fixed points for $f_\alpha(x,\mu)$ with respect to $x$ are marked with red circles in figure 2, panels a-f.
The accumulation depends on the hydrological cycle, sea ice cover and snow to rain ratio. Furthermore,
it could depend on the temperature of the deep ocean through the sea ice shift (SIS) mechanism suggested
by Gildor and Tziperman \cite{gildor:2000,tziperman:2003}. This is a good candidate for
explaining the difference between the early Pleistocene (right panels) and the late Pleistocene
(left panels) climates. The major difference is seen by comparing panels b and e in figure 2.
Panel b represents the climate drift in the late Pleistocene period where the deep ocean
is cold \cite{ruddiman:1989}. The low temperature of the deep ocean effectively prevents heat 
exchange with the upper water masses, such that an extensive sea ice cover can develop \cite{tziperman:2003}. 
The barriere separating the states $G$ and $g$ could be caused by a change in
the hydrological cycle such that to the left of the barriere (towards state $G$) the accumulation 
decreases rapidly with temperature due to the growth of sea ice cover, while to the right
of the barriere (towards state $g$) the accumulation decrease with increasing temperature. This could
be due to the southward position of the summer polar front changing the precipitation pattern over
the ice sheets. Perhaps this would result in more rainfall and less snow fall over the ice-sheets
as temperature increases leading to the decreasing accumulation with temperature \cite{tziperman:2003}. 

However, the real climate has not, except perhaps for short transient periods, been
in such a state, thus it cannot be reconstructed from paleoclimatic evidence and we will have to rely
on physical reasoning and future realistic climate models for explanations. In the early Pleistocene 
period for the same forcing the barriere is absent (panel e).   

When the state $i$ is reached, the Antarctic ice core
records show that the level of atmospheric $CO_2$ has increased due to oceanic heating and
following $CO_2$ release. When the climate state $G$ is reached after two bifurcations
the atmospheric $CO_2$ concentration will gradually decrease as the ocean cools. Furthermore,
the large glaciers build up on land.

In the late Pleistocene period when the glacial state $G$ is reached the meridional heat transport
in the oceans is strongly reduced. This would trap the warm ocean waters in the tropics, leading to a slow heating of the 
deep tropical ocean. This would perhaps in turn lead to an oceanic climate more similar to the one in the early Pleistocene 
period, where the deep ocean was warmer \cite{tziperman:2003}. 
This could then imply the slow change of the position of the bifurcation point $\alpha$ in the direction
of its position during the early Pleistocene period. This would make
way for the orbital forcing to trigger the transition $G\rightarrow i$. The 
triggering of a transition is then a combination of the deterministic orbital 
forcing and the internal stochastic noise induced forcing. 

The forcing will trigger the transition
$G \rightarrow i$ way before the early Pleistocene value is reached. 
One could speculate that the slow heating of the tropical ocean during the state $G$ could also
be related to the lower
atmospheric $CO_2$ concentration in the late Pleistocene period \cite{maasch:1990,saltzman:1990}. 
However, one should stress that interpretation in terms of accumulation, ablation and ocean circulation
is suggestive. Only the bifurcation structure of the governing equation is obtained from
the paleclimatic records. Future realistic climate models with the same type of bifurcation structure
are necessary to substantiate these speculations.  

\section{Summary}                 
In summary the empirical stochastic model presented support the suggestion that the transition from the 
'41 kyr world' to the '100 kyr world' occurring approximately 1 Myr - 800 kyr BP 
is due to a structural change in the bifurcation diagram describing the stability of the system as
a function of the forcing. The glacial cycles are not solely a deterministic response to the
orbital cycles. The internal noise also plays a role in triggering the jumps 
between the different climatic states in the first part of the '100 kyr world', 
this makes the ice ages fundamentally unpredictable. 
Currently the state-of-the-art general circulation climate models are far from being able to simulate the
observed glacial climate variations. It is even not known if they possess a non-trivial bifurcation structure.
This lack of dynamical range might be due to underestimation of internal variability in too coarse resolution, thus
the climate noise is to weak to induce transitions from one stable climate state to another .
The identification of the dynamical bifurcation diagram from observations, should be a guideline
for identification of physical mechanisms and ultimately for building realistic glacial climate models.    

\section*{Appendix: Parametrizing the bifurcation diagram}
The fixed point curves in the hysteresis diagram determines the drift function $f_\alpha(x,\mu)$. They
are specified from parabolic curves: 
$$(x^j(\mu)-f^j)=(\mu - \lambda_i)^2 
\Rightarrow x^j(\mu)=f^j\pm \sqrt{\mu -\lambda^j},$$
thus the bifurcation point is specified by the cartesian coordinates $(\lambda^j,f^j)$. For $\mu < \lambda$ there are two complex conjugate roots
to the equation not corresponding to real fixed points. These are indicated by dashed lines in figure 1 and figure 6. For 
$\mu > \lambda^j$ there are two branches (a stable and an unstable) of fixed points. The branches corresponding to
the intermediate glacial state 'g' and an unstable branch are thus specified as $$x_g^\pm(\mu) = f_g \pm \sqrt{\lambda_g - \mu},$$
where the bifurcation point is $(\lambda_g,f_g)=(0.1,-3.0)$,
see figure 6. Likewise there are two branches for the interglacial state: $$x_i^\pm(\mu) = f_i \pm \sqrt{\lambda_i - \mu},$$
with $(\lambda_i,f_i)=(-0.3,-0.5)$ and for the deep glacial state: $$x_G^\pm(\mu) = f_G \pm \sqrt{\lambda_G - \mu},$$
with $(\lambda_G,f_G)=(\alpha(t),-4.0)$. The parameter $\alpha(t)$ thus determines the horizontal position of the 
lower bifurcation point in figure 6. Now, as seen in figure 6, only the upper (stable) branch, $x_i^+$, and the lower (stable)
branch, $x_G^-$, are plotted. The curve connecting the two bifurcation points is simply given as a linear interpolation
between the two unstable branches: $$x_{iG}=(1-\lambda)x_i^- +\lambda x_G^+,$$
where $\lambda=(\mu-\lambda_G)/(\lambda_i-\lambda_G)$ and $\lambda_i\le \mu\le\lambda_G$. By this the five fixed points are now
defined and the drift function is $$
f_\alpha(x,\mu)=-(x-x_g^+)(x-x_g^-)(x-x_i^+)(x-x_G^-)(x-x_{iG}).$$ The first minus sign ensures that $f > 0$ for $x\rightarrow -\infty$ and
$f<0$ for $x\rightarrow \infty$ as it should. This is the form specified in equation \ref{poly}.
The parameter $\alpha(t)$ is specified from the climate state. If the climate is in the interglacial state the parameter 
is set to $\alpha=\alpha_1$. When the climate jumps to the glacial state 'G', the parameter change by an exponential
decay to $\alpha_0$ with timescale $\tau$: 
$$\frac{d(\alpha -\alpha_0)}{dt}=-\frac{(\alpha -\alpha_0)}{\tau}.$$ 
The parameters used here are $(\tau, \alpha_0, \alpha_1)=(10^5 yr, 0.5, 2.2)$.  

{\bf Acknowledgement: }This research was supported in part by the Nationl Science Foundation under Grant No.
PHY05-51164. 

\newpage
\bibliographystyle{authordate1}
\bibliography{/disk3/pditlev/documents/manus/climate}     

\newpage
\begin{center}
FIGURE CAPTIONS
\end{center}
\newcounter{fig}
\begin{list}{Fig. \arabic{fig}}
{\usecounter{fig}\setlength{\labelwidth}{2cm}\setlength{\labelsep}{3mm}}

\item
The bifurcation diagram for the model. Along the x-axis is the forcing represented
by the control parameter $\mu$, along the y-axis are the fixed points $\{x_0(\mu)|f(x_0,\mu)=0\}$ of the 
drift function $f_\alpha (x, \mu)$. The drift function is simply approximated by a
fifth order polynomial, with the roots determined by the fixed points. The horizontal
dashed line-segments indicates (real part of) sets of complex conjugate roots.
The fat curves show the stable fixed points. The
bifurcation point $\alpha$ is the point where the deep glacial state $G$ disappears. 
The arrows indicate the hysteresis loop as the forcing parameter is changed. Upper panel: The
glacial state $G$ is not accessible.
Lower panel: Now the location of the bifurcation point $\alpha$ has changed
in such a way that the deep glacial state $G$ is accessible. 

\item
The top panel shows a realization of the model. Second panel shows the orbital forcing 
driving the model. The red curve shows $\alpha(t)$, where the jumps to $\alpha=\alpha_1$ are
triggered by the transition $G \rightarrow i$. The next transition is in the low noise limit 
only possible when the blue curve is above the red curve. The third panel shows a "pseudo paleorecord", where a red noise component 
representing the non-climatic noise, is added to the model realization in the top panel. 
Lower panel shows the (normalized) paleoclimatic isotope record from a composite of ocean cores. 
The record is a proxy for the global sea level or minus the 
global ice volume. See text for more explanations.

\item
Five realizations of the model with the same orbital forcing and different stochastic forcing. The
first panel shows a realization without stochastic forcing. This is the purely deterministic climate
response to the orbital forcing. The bottom four realizations has a noise intensity $\sigma=0.8$ K/$\sqrt{kyr}$.
It is seen that only in the last part of the 100 kyr world the timing of the terminations are independent
from the noise. In all five realizations an additional non-climatic "proxy noise" is added aposteriory.  

\item
Upper panel: The spectral power of the sediment records for the two periods, 0-1 Myr and 1-2 Myr.
The red markers indicate 100 kyr, 41 kyr and 23 kyr periods.
The lower panels show the corresponding spectra for the randomly chosen realization of the model shown in figure 3.
Both the climate record and the model show a transition from the '41 kyr world' to the '100 kyr world'.  

\item
The form of the drift function as a function of the orbital forcing. The panels a-c corresponds
to the vertical intersections in the top left panel. The panels d-f are the intersections in the
top right panel. The major difference between the late Pleistocene and early Pleistocene periods
is seen by comparing panels b and e corresponding to the same orbital forcing in the two periods.
When the climate is in the glacial state G, there will be a slow lowering of the barrier separating
$G$ and $g$ indicated by arrows in top left panel and panel b. The (arbitrary) splitting of the 
drift function into an ablation part and an accumulation part is only suggestive.  

\item
The bifurcation diagram with stable branches $x_i^+$, $x_g^+$ and $x_G^-$. The three bifurcation 
points are $(\mu, x(\mu))=(\lambda_i, f_i)$ where the interglacial state disappears, $(\lambda_g, f_g)$ where the intermediate glacial state disappears and finally $(\lambda_G, f_G)$ where the
deep glacal state disappears. See text for explanation. 
\end{list}

\newpage
\begin{figure}[!H]
\begin{center}\epsfxsize=12cm 
\epsffile{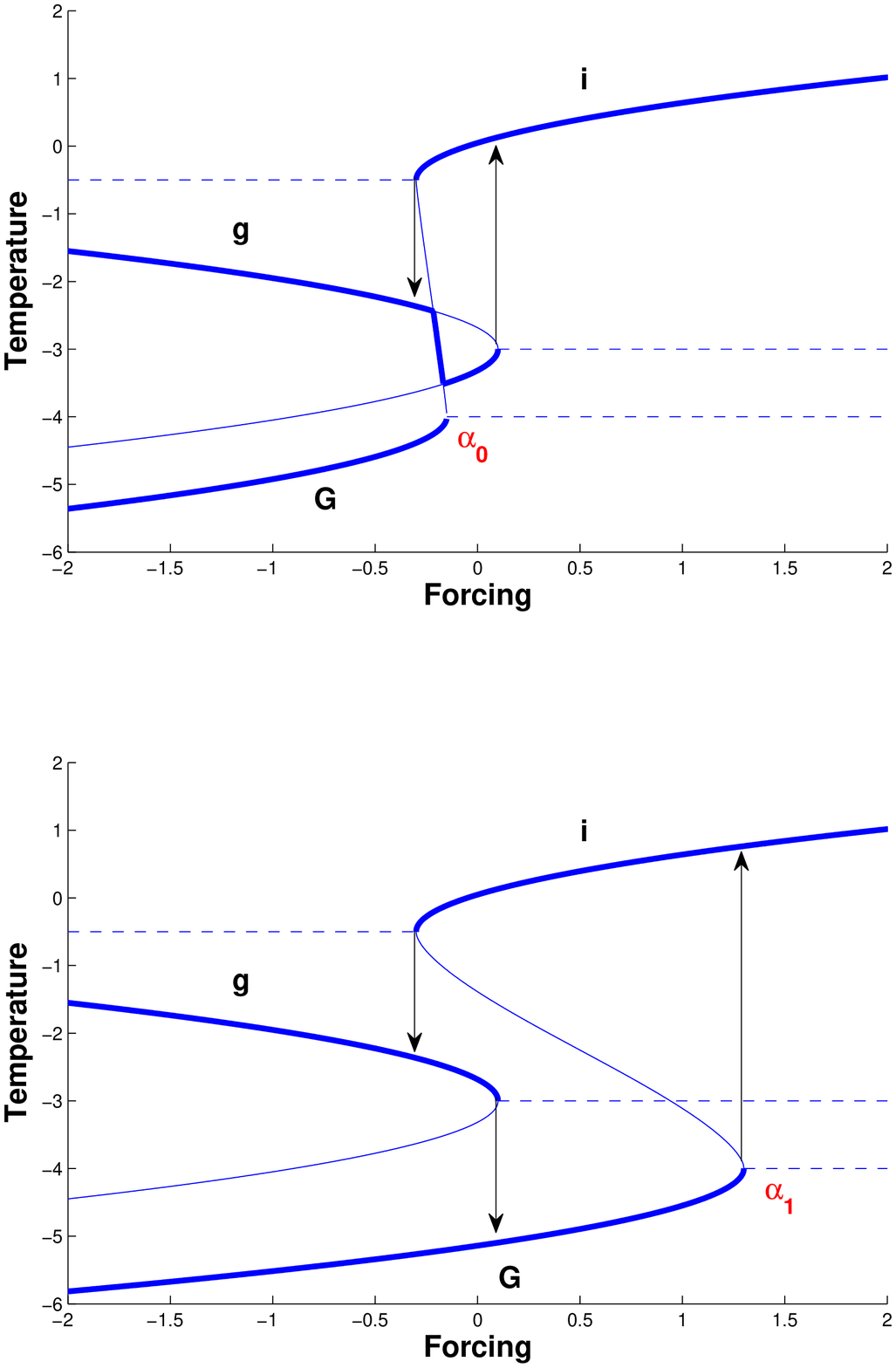}\caption[]{}
\end{center}\end{figure}

\newpage
\begin{figure}[!H]
\begin{center}\epsfxsize=\textwidth
\epsffile{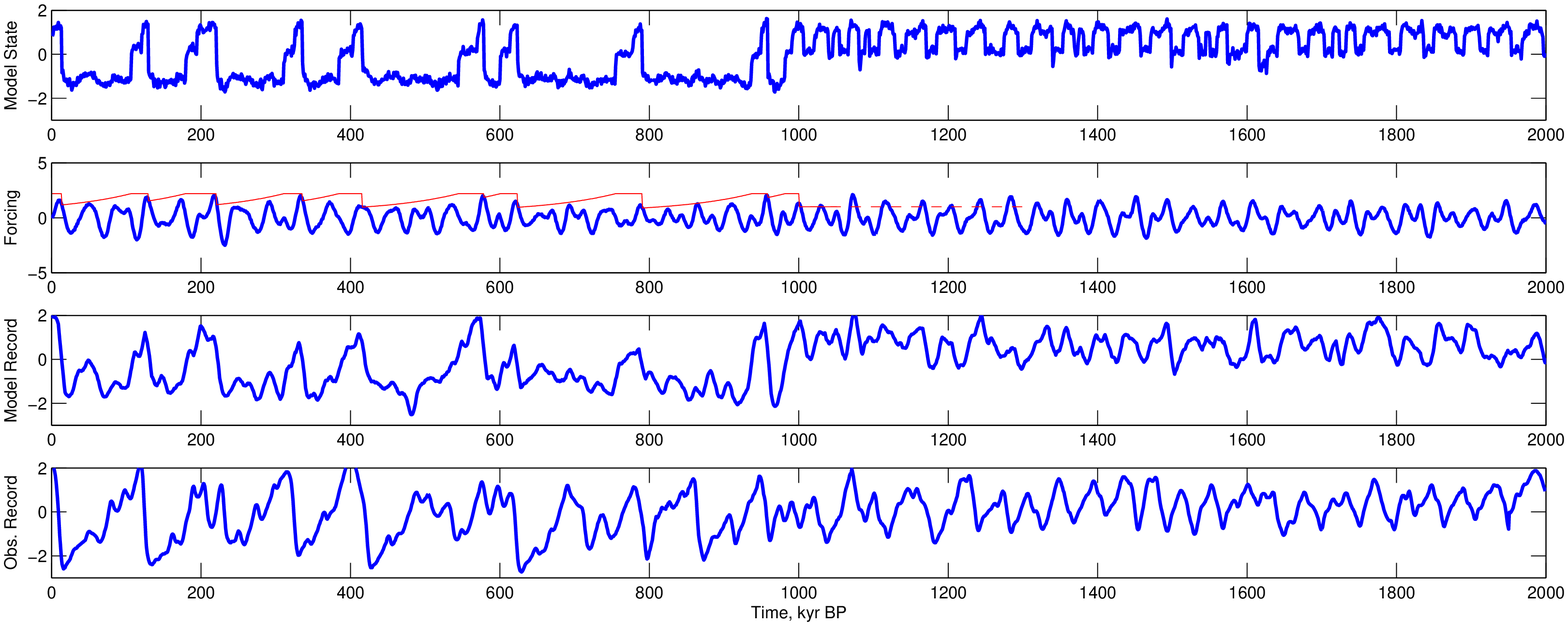}\caption[]{}
\end{center}\end{figure}

\newpage
\begin{figure}[!H]
\begin{center}
\epsfxsize=\textwidth
\epsffile{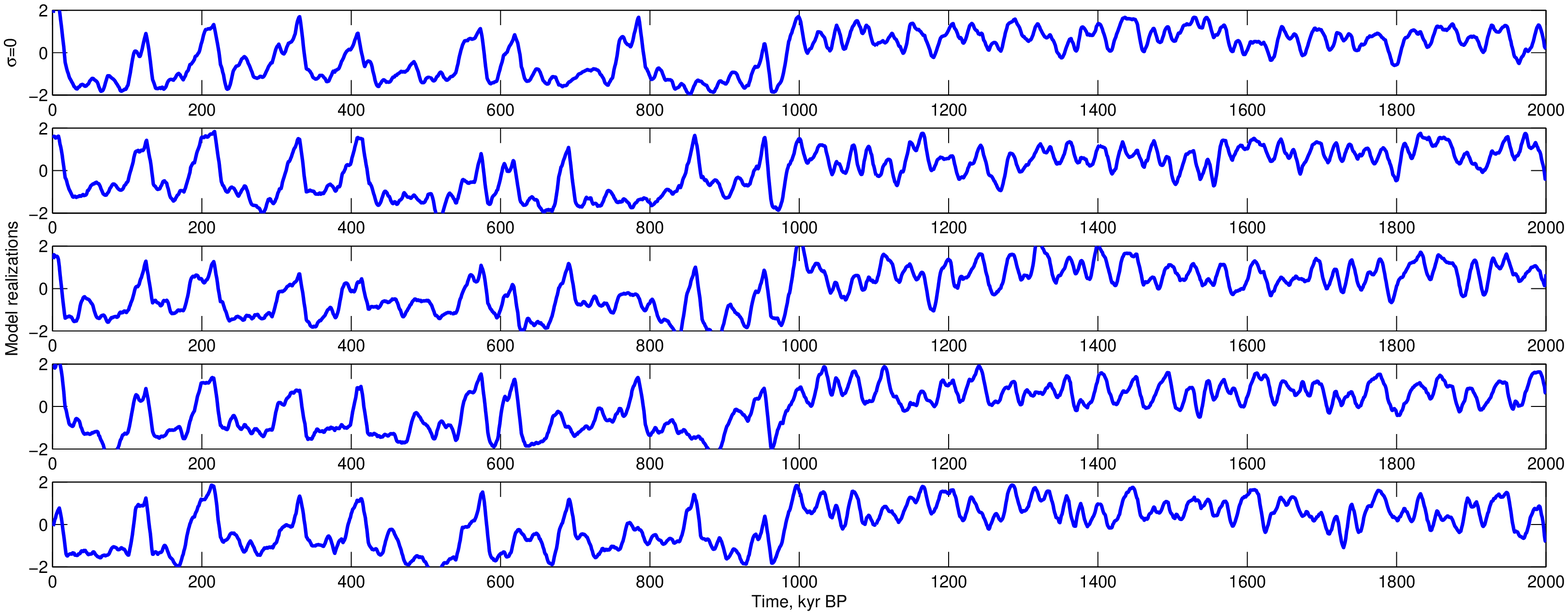}\caption[]{}
\end{center}
\end{figure}

\newpage
\begin{figure}[!H]
\begin{center}\epsfxsize=12cm 
\epsffile{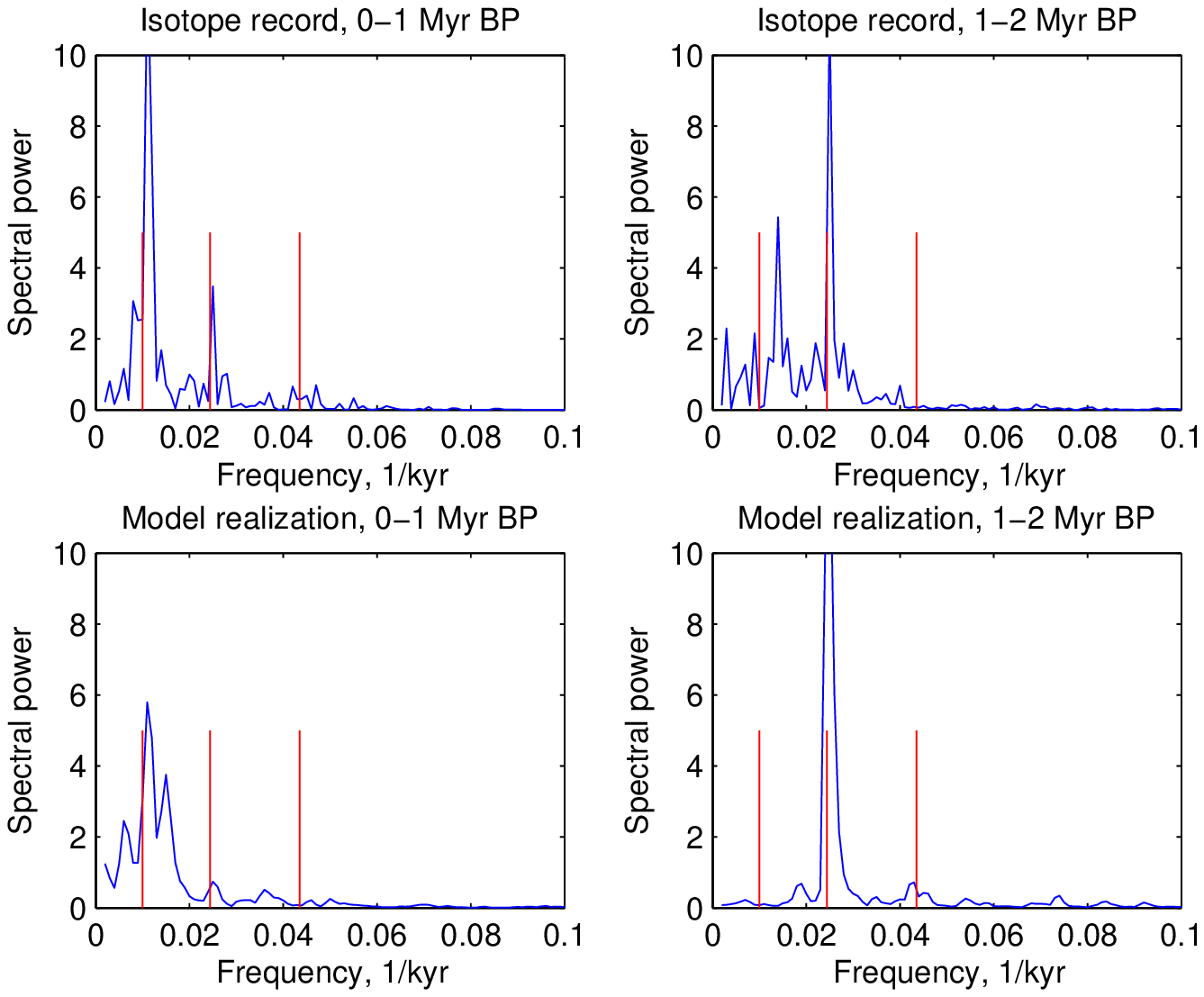}\caption[]{}
\end{center}\end{figure}

\newpage
\begin{figure}[!H]
\begin{center}\epsfxsize=12cm 
\epsffile{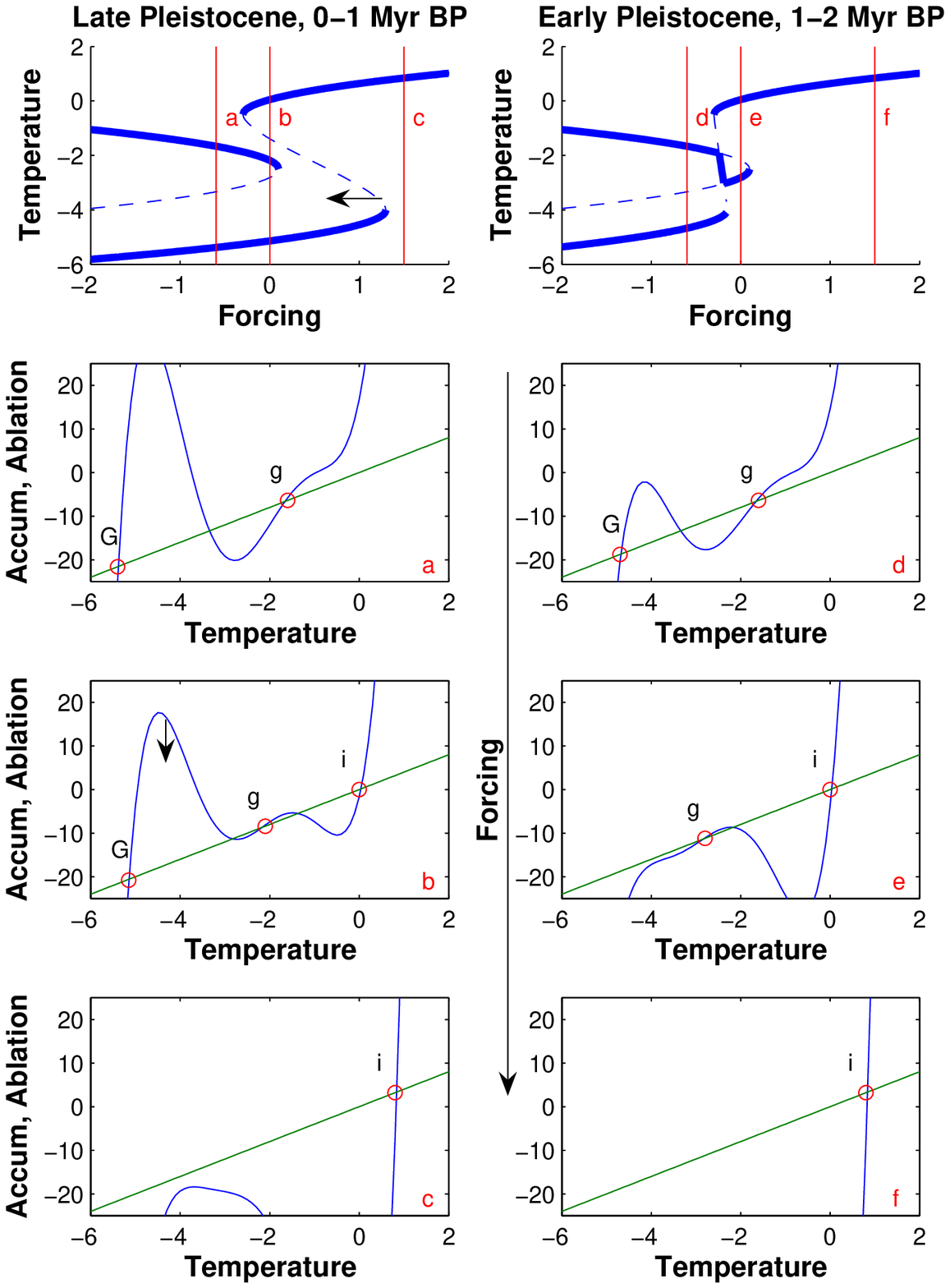}\caption[]{}
\end{center}\end{figure}

\newpage
\begin{figure}[!H]
\begin{center}\epsfxsize=12cm 
\epsffile{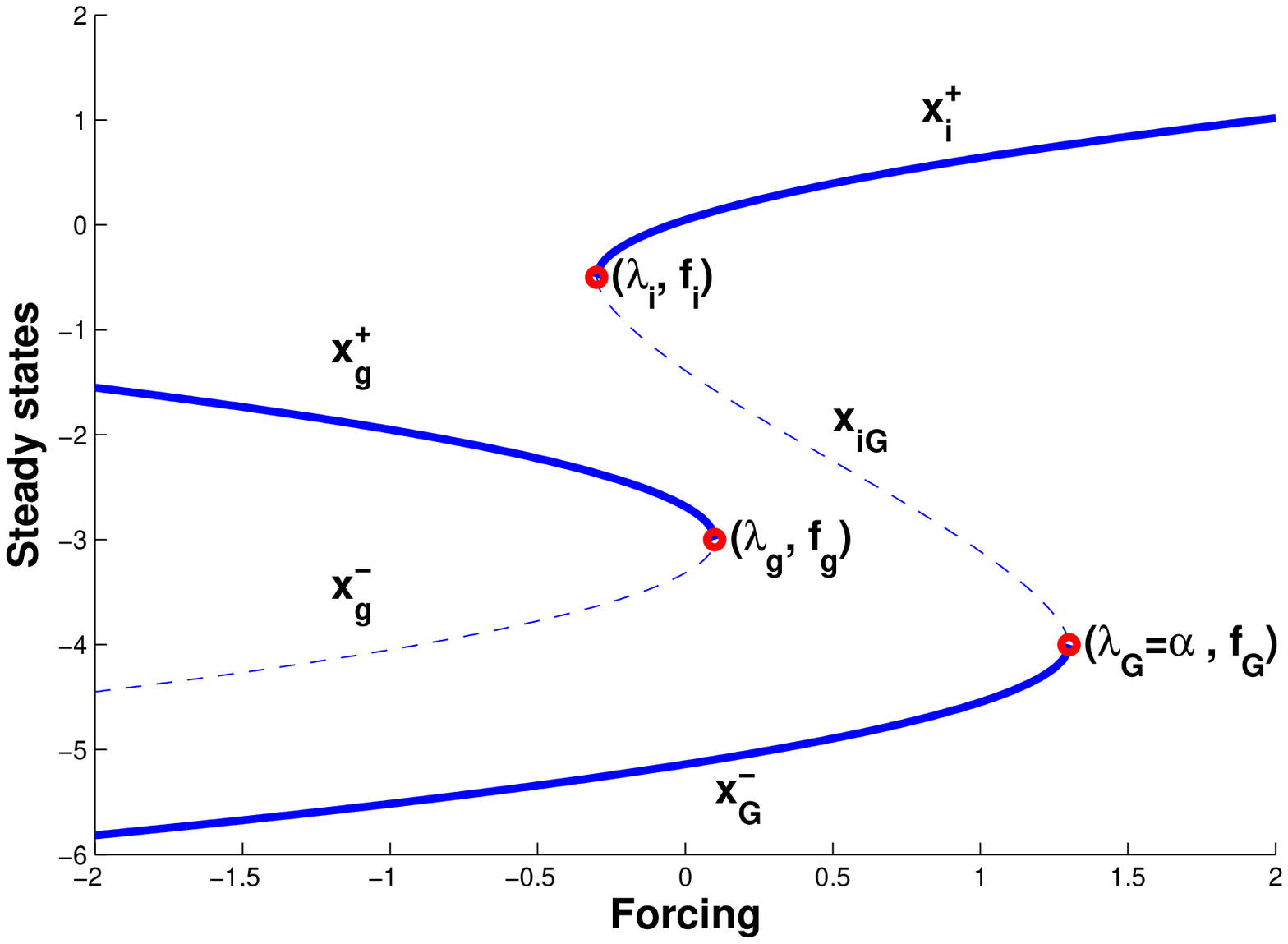}\caption[]{}
\end{center}\end{figure}
\end{document}